\documentclass[12pt]{article}
\usepackage{amsthm,amsfonts,amsmath,amssymb,mathrsfs,tikz,url,hyperref}
\usepackage{footmisc}
\usepackage{authblk}
\usepackage{mathtools}
\usepackage{physics}
\usepackage{dsfont}
\usepackage{float}
\usepackage[normalem]{ulem} 

\title{Dirac Wave Functions of Positive Energy with Arbitrarily Small Position Uncertainty}

\author{Ilmar B\"urck\footnote{Fachbereich Mathematik, Eberhard-Karls-Universit\"at T\"ubingen, Auf der Morgenstelle 10, 72076 T\"ubingen, Germany. E-mail: ilmar.buerck@student.uni-tuebingen.de} \, and
Roderich Tumulka\footnote{Fachbereich Mathematik, Eberhard-Karls-Universit\"at T\"ubingen, Auf der Morgenstelle 10, 72076 T\"ubingen, Germany. E-mail: roderich.tumulka@uni-tuebingen.de }}
\date{July 2, 2026}

\newcommand{\be}{\begin{equation}}
	\newcommand{\ee}{\end{equation}}
\newcommand{\Hilbert}{\mathcal{H}}

\newcommand{\ve}{\boldsymbol{e}}
\newcommand{\vp}{\boldsymbol{p}}
\newcommand{\vx}{\boldsymbol{x}}
\newcommand{\vzero}{\boldsymbol{0}}
\newcommand{\RRR}{\mathbb{R}}

\theoremstyle{plain}
\newtheorem{thm}{Theorem}
\newtheorem{lem}{Lemma}
\theoremstyle{definition}

\newtheorem{remark}{Remark}

\begin{document}
\maketitle
\begin{abstract}
We consider wave functions in the Hilbert space $\Hilbert=L^2(\mathbb{R}^3,\mathbb{C}^4)$ of a single Dirac particle, specifically from the positive-energy subspace $\Hilbert_+$ of the free Dirac Hamiltonian. Over the decades, various authors hypothesized that for wave functions from $\Hilbert_+$, there is a positive lower bound to the position uncertainty $\sigma_x$; in other words, that such states cannot be arbitrarily narrow in $x$. Using a sequence of wave functions introduced by Bracken and Melloy, we show that this hypothesis is false. (In fact, they already stated that it is false, but their proof that their sequence is a counter-example had a gap.)

\medskip
		
\end{abstract}

\section{Introduction}
\label{Introduction}

There are many questions surrounding the possibility of localizing a relativistic quantum particle; {some recent works in this direction include \cite{Cho24,DRM24,DM25,Tu26,Mor26a,Mor26b}}. We consider here a single particle with wave function $\psi$ governed by the Dirac equation,
\begin{align}
    i\frac{\partial}{\partial t}\ket{\psi} = H_D \ket{\psi}.
\end{align}
Here, $\ket{\psi}$ is an element of the Hilbert space $\Hilbert=L^2(\mathbb R^3, \mathbb C^4)$, which is the space of square integrable $\mathbb{C}^4$ valued functions. $H_D$ is the free Dirac Hamiltonian, which in natural units ($\hbar=1=c$) is given by
\begin{align}
    H_D = -i\boldsymbol{\alpha}\cdot \nabla + \beta m\,,
\end{align}
where $\boldsymbol \alpha=(\alpha_1,\alpha_2,\alpha_3)$ and $\beta$ are the Dirac matrices {and $m>0$ is the mass}. The spectrum of $H_D$ is $(-\infty,-m]\cup [m,\infty)$ and thus, $\Hilbert$ is the orthogonal sum of the positive and negative energy subspaces $\mathcal{H_{\pm}}$. As negative energies are usually considered unphysical for a free electron, we focus on vectors in $\Hilbert_+$.

There are two facts that clash with the localizability of such particles. The first is that any generalized eigenfunction of the position operator, $\chi \, \delta^3(\vx)$ with arbitrary spinor $\chi\in \mathbb{C}^4$, has equally strong contributions of positive and negative energy, so it cannot be associated with only $\Hilbert_+$.
Secondly, there exist no wave functions in $\Hilbert_+$ that strictly vanish outside a bounded subset of $\mathbb{R}^3$. Even more, for any $\psi\in\Hilbert_+$ with $\psi\neq 0$, the set $\{\vx \in\mathbb{R}^3| \psi(\vx) \neq 0\}$ is dense in $\mathbb{R}^3$ \cite[Cor.~1.7]{Thaller_1992} (so that its closure, known as the support of $\psi$, is all of $\mathbb{R}^3$). In particular, any $\psi\in\Hilbert_+$ has tails reaching to spatial infinity.
This is surprising because one might have expected that a detection of the particle in some bounded set $B\subset\mathbb{R}^3$ would collapse the wave function so that it vanishes outside $B$.

In this paper, we consider another aspect of localization that can still be applied to functions with tails to infinity: the position uncertainty $\sigma_x$ of wave functions $\psi\in\Hilbert_+$ with $\|\psi\|=1$, which we will henceforth denote as $\sigma_\psi$ (and also refer to as the spread or width or size of $\psi$). It is defined as the standard deviation of the Born distribution for position, which has density $\rho_\psi(\vx) = |\psi(\vx)|^2 = \psi(\vx)^\dagger \, \psi(\vx)$. Equivalently, $\sigma_\psi\geq 0$ is defined by
\begin{align}
    \sigma_\psi^2 
    &= \int d^3\boldsymbol x\: \boldsymbol x^2|\psi(\boldsymbol x)|^2 - \left(\int d^3\boldsymbol x \: \boldsymbol x |\psi(\boldsymbol x)|^2 \right)^2\\
    &= \langle \psi|\vx^2|\psi\rangle - \langle \psi|\vx|\psi\rangle^2\,.
\end{align}
Various authors \cite{Hei3rd,RevModPhys.30.24,BD64,Sakurai,RevModPhys.40.632, 10.1119/1.13795,CCN10,Seb19}, \cite[Sec.~4.1]{GTZ24} have hypothesized or even claimed that there is a positive lower bound for $\sigma_\psi$ with $\psi\in\Hilbert_+$, meaning that such $\psi$ cannot be arbitrarily narrow; see \cite{Sebens_2020} for an overview. While there are heuristic arguments in support of this claim as we explain in Section~\ref{sec:arguments}, we show here that this claim is wrong:

\begin{thm}\label{thm:1}
For $\psi\in\Hilbert_+$ with $\|\psi\|=1$, $\sigma_\psi$ can be arbitrarily close to 0. In other words, for every $\varepsilon>0$ there exists $\psi\in\Hilbert_+$ with $\|\psi\|=1$ and $\sigma_\psi < \varepsilon$. In other words, there is a sequence $\psi_n\in\Hilbert_+$ with $\|\psi_n\|=1$ such that $\sigma_{\psi_n}\to 0 $ as $n\to\infty$. 
\end{thm}

(All proofs are collected in Section~\ref{sec:proofs}.)

This fact has already been stated by Bracken and Melloy \cite{Bracken_1999,Mel02}, and in fact they have constructed in \cite{Bracken_1999} a sequence $\psi_n$ of the kind described in the last sentence of Theorem~\ref{thm:1}, but their proof that their sequence has the desired properties had a gap that we close here. Indeed, they showed of their sequence $\psi_n$ that
\begin{align}
    \lim_{n\to\infty} \rho_{\psi_n}(\boldsymbol x) = \delta^3(\boldsymbol x) 
\end{align}
(the 3d delta function), and then assumed that if the probability densities approach a delta function, then the position uncertainty must tend to 0. However, this is not the case:

\begin{thm}\label{thm:2}
In any dimension $d\geq 1$, there exists a sequence of (measurable, even smooth) functions $\rho_n(\vx)\geq 0$ on $\mathbb{R}^d$ that satisfy
\begin{align}
    \int_{\mathbb{R}^d} d^d\vx \,  \rho_n(\vx) &= 1\\
    \lim_{n\to\infty} \rho_n(\vx) &= \delta^d(\vx)
    \label{eqn: delta property}\\
    \int d^d\vx\, \rho_n(\vx) \, \vx  &= \boldsymbol{0}
    \label{eqn: zero expectation value}\\
    \lim_{n\to\infty}\int_{\mathbb{R}}d^d\vx \, \rho_n(\vx) \, \vx^2 &= \infty\,.
    \label{eqn: infinite variance}
\end{align}
\end{thm}

That is, if a sequence of probability densities converges to $\delta^d(\vx)$, then, although that is some sense in which the distribution becomes narrow (e.g., it implies that for any compact $K\subset \mathbb{R}^d$ not containing the origin, the probability content $\int_K d^d\vx \, \rho_n(\vx)$ tends to 0 as $n\to\infty$), it is not sufficient to ensure that the standard deviation tends to 0.

\begin{remark}\label{rem:converse}
While the convergence of a sequence $\rho_n(\vx)$ to the delta function does not guarantee that the variance is 0, the reverse claim holds. That is,
\begin{equation}
    \text{if }\lim_{n\to\infty}\int d^d\vx \, \vx^2 \rho_n(\vx) = 0\,, \text{ then }
    \lim_{n\to\infty}\rho_n(\vx) = \delta^d(\vx).
\end{equation}
\hfill$\diamond$
\end{remark}

Our proof of Theorem~\ref{thm:1} does not involve delta functions, and indeed is simpler than the proof that $\rho_{\psi_n}\to\delta^3$. Rather, the proof directly provides an upper bound on $\sigma_{\psi_n}$.

\bigskip

The remainder of this note is organized as follows. In Section~\ref{sec:arguments}, we elucidate why one might have expected a positive minimal position uncertainty. In Section~\ref{sec:construction}, we recall the construction of the sequence $\psi_n$ used in the proof of Theorem~\ref{thm:1}. In Section~\ref{sec:proofs}, we present the proofs. {In Section~\ref{sec:literature}, we compare our result to related ones in the literature.}

\section{Claims and Arguments for a Positive Minimal Size}
\label{sec:arguments}

Although false, the idea of a positive minimum position uncertainty is not implausible, as some heuristic arguments in its favor exist, which we outline now. (See also \cite{Sebens_2020} for further discussion.)

\subsection{Pair Production}

When $\sigma_x$ is much smaller than the Compton wave length
\begin{equation}\label{lambdaC}
\lambda_C=\frac{2\pi\hbar}{mc} \,,
\end{equation}
then the Heisenberg uncertainty relation should entail that the momentum uncertainty $\sigma_p$ is much larger than $\hbar/2\lambda_C = mc/4\pi$, and thus that much of the energy available is greater than $2mc^2$, which would be enough for the production of an electron--positron pair. While this does not prove any mathematical relation, it seems to indicate a physical reason why it might be impossible to make $\psi\in\Hilbert_+$ much narrower than $\lambda_C$ \cite[p.~193]{Hei2nd}: because when we try to do this, the quantum state will not remain a 1-particle state.\footnote{{This is related to the fact that in quantum electrodynamics, 1-electron states of negative energy get transformed into 1-positron states of positive energy (e.g., \cite[Ch.~10]{Thaller_1992}). Furthermore, it seems correct that if we physically force an electron to have $\sigma_x\ll \lambda_C$, then pair production should occur, so that the 1-particle Dirac equation is no longer applicable; but this is distinct from the mathematical fact expressed by Theorem~\ref{thm:1}.}}

\subsection{Width of $P_+\delta^3$}

\begin{figure}[ht]
  \centering
  \includegraphics[width=0.6\linewidth]{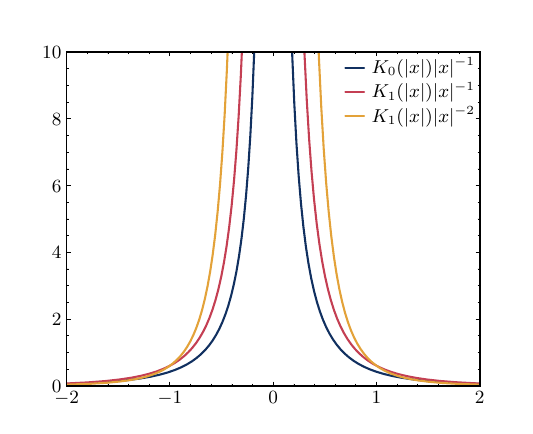}
  \caption{Plots of the functions $K_0(|x|)/|x|$, $K_1(|x|)/|x|$ and {$K_1(|x|)/|x|^2$} that show up as coefficients in \eqref{eqn: Kernel}}
  \label{fig:Bessel}
\end{figure}

We write $\widehat{\psi}(\boldsymbol{p})$ for the Fourier transform of $\psi(\vx)$. Consider the projection operator $P_+$ onto the positive energy subspace $\Hilbert_+$. In momentum space, this operator can be written as a matrix-valued multiplication operator $P_+(\boldsymbol p)$, meaning that for an arbitrary wave function $\psi\in\Hilbert$ we have that
\begin{align}
    \widehat{(P_+ \psi)}(\boldsymbol p) = P_+(\boldsymbol p) \, \widehat{\psi}(\boldsymbol p) \,.
\end{align}
By the convolution theorem, this means that in position space
\begin{align}
    (P_+\psi)(\boldsymbol{x}) =  \frac{1}{(2\pi)^\frac{3}{2}}\int_{\mathbb{R}^3} d^3\boldsymbol{y}\, \check{P}_+(\boldsymbol x - \boldsymbol y) \, \psi(\boldsymbol y),
    \label{eqn: convolution}
\end{align}
where $\check{P}_+(\vx)$ is the inverse Fourier transform of the matrix-valued function $P_+(\boldsymbol{p})$, given by 
\begin{align}
    \check P_+(\boldsymbol x) &= \beta\frac{m^2}{(2\pi)^\frac{1}{2}|\vx|}K_1(m|\vx|) \: +\nonumber\\
    &+ \boldsymbol \alpha \cdot \boldsymbol x \left(i\frac{m^2}{(2\pi)^\frac{1}{2}|\vx|^2}K_0(m|\vx|) + i \frac{2m}{(2 \pi)^{\frac{1}{2}}|\vx|^3}K_1(m|\vx|)\right) + \frac{1}{2}\delta^3 (\boldsymbol x) \mathds{1}.
    \label{eqn: Kernel}
\end{align}
Here, $K_\nu$ denote the modified Bessel functions of the second kind and $\mathds{1}$ the unit matrix. A derivation of this expression for $\check P(\boldsymbol x)$ can be found in Section~\ref{sec:kernel}. Now, consider a state that is localized at 0, that is,
\begin{align}\label{psideltadef}
    \psi(\boldsymbol x) = \delta^3( \boldsymbol x) \, e_s \,,
\end{align}
where $e_1,\ldots,e_4$ are the canonical basis vectors of $\mathbb C^4$. Although $\psi$ is, due to its delta function, not an element of $\Hilbert$, it is formally possible to apply $P_+$ to it by means of \eqref{eqn: convolution}, which yields that
\begin{align}
    (P_+\psi)(\boldsymbol x) = \frac{1}{(2\pi)^\frac{3}{2}}\check{P}_+(\boldsymbol x) \, e_s \,,
\end{align}
which is proportional to the $s$-th column of the matrix $\check{P}_+(\vx)$. That is, we can give meaning to the positive-energy part of $\delta^3$, and it turns out to be a function with 4 contributions visible in \eqref{eqn: Kernel}: three Bessel functions depicted in Figure~\ref{fig:Bessel} and a delta function.

It is apparent from the figure that the functions $K_\nu(|x|)/|x|$ and $K_1(|x|)/|x|^2$ have a certain positive width of order 1 (although $K_\nu(|x|)/|x|$ is actually not square integrable, which poses an obstacle to making this consideration quantitative). This suggests that $K_\nu(m|\vx|)/m|\vx|$ and $K_1(m|\vx|)/m^2|\vx|^2$ have width of order $1/m$, which means (in view of our convention $\hbar=1=c$) a width of order of the Compton wave length $\lambda_C$ \eqref{lambdaC}. 
The fact that even a delta function, when projected to the positive-energy subspace, has width of order $\lambda_C$, might suggest that any function from $\Hilbert$, when projected to $\Hilbert_+$, should have width at least as much.

(As a similar consideration, the eigenfunctions of the Newton-Wigner position operator 
\cite{NW,Thaller_1992,Beck}, expressed in the standard position representation, have apparent positive width, which might suggest again that their width is the minimum width of $\psi\in\Hilbert_+$ \cite{RevModPhys.30.24}, \cite[p.~119]{Sakurai}.)

\subsection{Convolution Argument}

For two positive normalized functions $f$ and $g$, their convolution $f*g$ will have a larger standard deviation than the individual functions; that is, when
\begin{align}
    \int dx \, f(x) = \int dx \, g(x) = 1,\\
    (f*g)(x) = \int dy \, f(x-y) \, g(y),
\end{align}
then
\begin{align}\label{sigma2add}
    \sigma^2_{f*g} = \sigma^2_f + \sigma^2_g \,.
\end{align}
This fact is well known in probability theory because $f*g$ is the probability density of $X_f+X_g$ if $X_f$ and $X_g$ are independent random variables with probability densities $f$ and $g$, and the variance of a sum of independent random variables is the sum of the variances.

Since for any $\Psi\in\Hilbert$ its projection $P_+\Psi$ is, by \eqref{eqn: convolution}, proportional to the 
convolution $\check{P}_+ * \Psi$, and $\check{P}_+$ has position spread of order $\lambda_C$, it is tempting to assume that all wave functions in the positive energy subspace (i.e., resulting from such projection) must have position spread at least of order $\lambda_C$. 

A reasoning of this kind does actually apply in another situation: to positive solutions $u(t,\vx)$ of the heat equation
\be
\partial_t u= \nabla^2 u
\ee
with initial datum $u_0(\vx)$; the solution is known \cite{heat} to be given by
\be
u_t = g_t * u_0 \text{ with } g_t(\vx) = (4\pi t)^{-3/2} \exp(-\vx^2/4t) \,.
\ee
Indeed, if $u_0\geq 0$ and $\int d\vx\, u_0(\vx)=1$, then $\sigma^2_{u_t}=\sigma^2_{u_0}+t/2>\sigma^2_{u_0}$ by \eqref{sigma2add}.

Of course, the convolution reasoning presupposes that the functions involved are non-negative, whereas $\Psi$ is not so constrained and in fact spinor-valued; in other words, cancellations can undermine the increase in the spread. That is why the reasoning is not compelling in our case.

\section{Review of the Bracken-Melloy Construction}
\label{sec:construction}

The sequence $\psi_n$ is constructed using an arbitrary fixed function $f(\boldsymbol p)$, for example a Gaussian, that is normalized, i.e.,
\begin{align}
    \int_{\mathbb R ^3}d^3\boldsymbol p \: |f(\boldsymbol p)|^2 = 1.
    \label{eqn: normalization of f}
\end{align}
Bracken and Melloy required $f(\boldsymbol{p})$ to be a Schwartz function, but our proof still works with the weaker requirement (which Schwartz functions satisfy\footnote{Recall that Schwartz functions are defined by the property that for any integer $N$ and any multi-index $\boldsymbol \gamma$,  $\sup_{\boldsymbol p}|\boldsymbol{p}|^N |\partial^{\boldsymbol{\gamma}} f(\boldsymbol{p})|<\infty$. From this it follows that the first integral of \eqref{eqn: conditions} is finite, as possible tails to infinity decay rapidly and the integrand is bounded. By the same reason, the tails of the second and third integral pose no problem, which leaves us with the pole at the origin. By switching to spherical coordinates, one sees that the $p^2$ factor in the volume element dominates the $1/|\boldsymbol p|$ and $1/|\boldsymbol p|^2$ factors, ensuring that the integral around the origin is finite.})
that the following integrals are all finite:
\begin{align}
    \int_{\mathbb{R}^3}d^3\boldsymbol p \left|\frac{\partial f}{\partial p_j}\right|^2  <\infty,~~
    \int_{\mathbb{R}^3}d^3\boldsymbol p \left|\frac{\partial f}{\partial p_j}\frac{f(\boldsymbol p)}{|\boldsymbol p|}\right| <\infty,~~
    \int_{\mathbb{R}^3}d^3\boldsymbol p \frac{|f(\boldsymbol p)|^2}{\boldsymbol p^2} <\infty.
    \label{eqn: conditions}
\end{align}

Let $u(\boldsymbol p)$ be the momentum-dependent Dirac spinor given by
\begin{align}\label{udef}
    u(\boldsymbol p) = \frac{(E(\boldsymbol p)+m,0,p_3,p_1-ip_2)}{\sqrt{2E(\boldsymbol p)(E(\boldsymbol p)+m)}}.
\end{align}
This choice ensures
\begin{align}
    H(\boldsymbol p) \: u(\boldsymbol p) &= E(\boldsymbol p) \: u(\boldsymbol p),\\
    |u(\boldsymbol p)| &= 1,
    \label{eqn: spinor is normalized}
\end{align}
with
\begin{align}
    E(\boldsymbol p) &= \sqrt{\boldsymbol p^2+ m^2},\label{Edef}\\
    H(\boldsymbol p) &= \boldsymbol \alpha\cdot \boldsymbol p + \beta m.
\end{align}
The spinor $u(\boldsymbol{p})$ is known as a member of an orthonormal basis of $\mathbb{C}^4$ diagonalizing $H(\boldsymbol{p})$ (in \cite[(1.43)]{Thaller_1992}, $\mathrm{u}(\boldsymbol{p})$ denotes the matrix whose columns are this basis).

The sequence $\psi_n$ is now defined by
\begin{align}
    \widehat\psi_n(\boldsymbol p) &= \frac{1}{n^{3/2}} \:f\left(\frac{\boldsymbol p}{n}\right) \: u(\boldsymbol p),
    \label{eqn: sequence of wave functions in momentum space}\\
    \psi_n(\boldsymbol x) &= \frac{1}{(2\pi)^{3/2}}\int_{\mathbb R^3}d^3\boldsymbol p \: \widehat\psi_n(\boldsymbol p) \: e^{i\boldsymbol x\cdot \boldsymbol p} .
    \label{eqn: sequence of wave functions}
\end{align}
Using \eqref{eqn: normalization of f} and substituting a new variable for $\boldsymbol p/n$, we see that the wave functions are normalized,
\begin{align}
\int d^3 \boldsymbol x \: |\psi_n(\boldsymbol x)|^2 = \int d^3\boldsymbol p \: |\widehat \psi_n(\boldsymbol p)|^2 = 1.
\end{align}

\begin{figure}[ht]
\centering
  \includegraphics[width=0.6\linewidth]{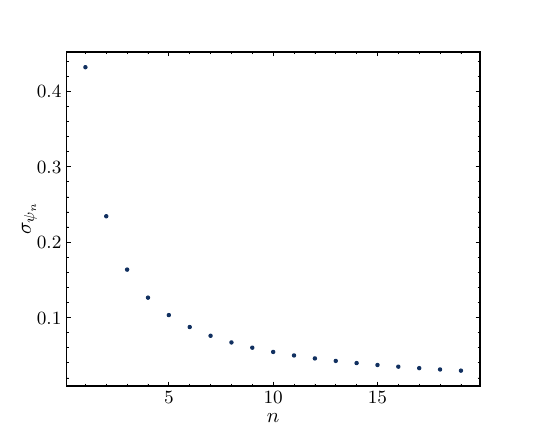}
  \caption{$\sigma_{\psi_n}$ for $n = 1,\dots,19$, illustrating that $\sigma_{\psi_n}\to 0$ as $n\to\infty$}
  \label{fig:test1}
\end{figure}

\begin{figure}[ht]
  \centering
  \includegraphics[width=0.6\linewidth]{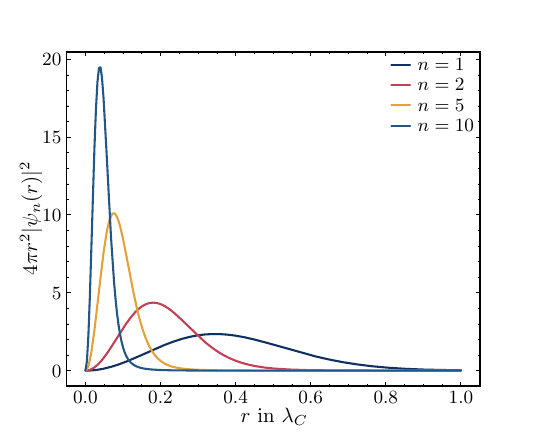}
  \caption{$4\pi r^2|\psi_n(r)|^2$ for $n$ = 1, 2, 5, 10 with $r$ in units of the Compton wave length}
  \label{fig:test2}
\end{figure}

Figures~\ref{fig:test1} and \ref{fig:test2} depict numerical calculations of the standard deviation $\sigma_{\psi_n}$ and the radial probability density of the sequence $\psi_n$ in agreement with Theorem~\ref{thm:1}.

\section{Proofs}
\label{sec:proofs}

\subsection{Derivation of the Kernel}
\label{sec:kernel}

The projection onto the positive energy subspace is given in momentum space by the multiplication operator
\begin{align}
    \widehat{P_+\psi}(p) &= \frac{H+E(\boldsymbol p )}{2 E(\boldsymbol p)} \widehat{\psi}(p)\\
    &= \underbrace{\left(\frac{\boldsymbol \alpha \cdot \boldsymbol p + \beta m}{2E(\boldsymbol p)} +\frac{1}{2}\mathds 1 \right)}_{ P_+(\boldsymbol p)} \widehat\psi(\boldsymbol p).
\end{align}
By the convolution theorem we have that
\begin{align}
    (P_+\psi)(\boldsymbol x) = \frac{1}{(2\pi)^{\frac{3}{2}}}\int d^3 \boldsymbol y \check{P}(\boldsymbol x-\boldsymbol y)\psi(\boldsymbol{y}).
\end{align}
Where $\check{P}(\boldsymbol{x})$ is the Fourier transform of $P_+(\boldsymbol{p})$. We have that
\begin{align}
    \check{P}(\boldsymbol{x}) &= \int \frac{d^3 \boldsymbol{p}}{(2\pi)^{\frac{3}{2}}}e^{i \boldsymbol{p} \cdot \boldsymbol{x}} P_+(\boldsymbol{p})\\
    &= \int \frac{d^3 \boldsymbol{p}}{(2\pi)^{\frac{3}{2}}}e^{i \boldsymbol{p} \cdot \boldsymbol{x}} \left(\frac{\boldsymbol \alpha \cdot \boldsymbol p + \beta m}{2E(\boldsymbol p)} +\frac{1}{2}\mathds 1 \right)\\
    &= \frac{1}{2}\beta m F(\boldsymbol{x}) + \frac{1}{2} \boldsymbol \alpha \cdot \boldsymbol G(\boldsymbol{x}) + \frac{1}{2}\delta^3 (\boldsymbol{x}) \mathds{1}.
\end{align}
Let $p = |\boldsymbol{p}|$ and $x = |\boldsymbol{x}|$. First we get
\begin{align}
    F(\boldsymbol{x}) &= \int \frac{d^3\boldsymbol{p}}{(2\pi)^{\frac{3}{2}}}\frac{e^{i\boldsymbol{p} \cdot \boldsymbol{x}}}{\sqrt{\boldsymbol{p}^2+m^2}}\\
    &=\frac{1}{(2\pi)^{\frac{3}{2}}}\int_0^\infty dp \int_{-\pi}^\pi d\varphi \int_0^\pi d\theta \frac{\sin(\theta)p^2e^{ipx \cos(\theta)}}{\sqrt{p^2+ m^2}}\\
    &= \frac{1}{(2\pi)^{\frac{1}{2}}}\int_0^\infty \int_{-1}^1 du\frac{p^2e^{ipxu}}{\sqrt{p^2 + m^2}}\\
    &=  \frac{1}{(2\pi)^{\frac{1}{2}}}\int_0^\infty dp \frac{2p\sin(px)}{x\sqrt{p^2+m^2}}\\
    &= -\frac{2}{(2\pi)^{\frac{1}{2}}x}\frac{d}{dx}\int_0^\infty dp \frac{\cos(px)}{\sqrt{p^2+m^2}}\\
    &= -\frac{2}{(2\pi)^{\frac{1}{2}} x} \frac{d}{dx}K_0(mx) \label{eqn: integral rep of bessel}\\
    &= \frac{2m}{(2\pi)^{\frac{1}{2}}x}K_1(mx)
    \label{eqn: F(x)}
\end{align}
Here, in step~\eqref{eqn: integral rep of bessel}, we used Basset's integral and in~\eqref{eqn: F(x)} we used that $K_0' = -K_1$,  see Eq.~10.32.11 in \cite{dlmf10_32} and \cite{dlmf10_29} respectively. 
We see that $F(\boldsymbol{x}) = F(x)$ only depends on the norm of $\boldsymbol{x}$. Consider the following identities that can be found in~\cite{dlmf10_29}:
\begin{align}
    K_1'(z) &= -\frac{1}{2}(K_0(z) + K_2(z))\\
    K_2(z) &= K_0(z) + \frac{2}{z}K_1(z)\,,
\end{align}
which results in
\begin{align}
    K_1'(z) = -K_0(z) - \frac{1}{z}K_1(z) \,.
    \label{eqn: K_1 derivative}
\end{align}
Using this, we find that
\begin{align}
    G_j(\boldsymbol{x}) &= \frac{1}{(2\pi)^{\frac{3}{2}}}\int d^3 \boldsymbol{p}\frac{p_j}{\sqrt{\boldsymbol{p}^2 + m^2}}e^{i \boldsymbol{p} \cdot \boldsymbol{x}}\\
    &= -i\frac{\partial}{\partial x_j}F(|\vx|)\\
    &= -i F'(|\vx|) \frac{x_j}{|\vx|}\\
    &= i\frac{2mx_j}{(2\pi)^{\frac{1}{2}}|\vx|^3}K_1(m|\vx|) - i\frac{2m^2x_j}{(2\pi)^{\frac{1}{2}} |\vx|^2}K_1'(m|\vx|)\\
    &\overset{\eqref{eqn: K_1 derivative}}{=}  i\frac{2m^2x_j}{(2\pi)^{\frac{1}{2}} |\vx|^2}K_0(m|\vx|) +i\frac{4mx_j}{(2\pi)^{\frac{1}{2}}|\vx|^3}K_1(m|\vx|).
    \label{eqn: G_j(x)}
\end{align}
Substituting the expressions in Eqs.~\eqref{eqn: F(x)} and \eqref{eqn: G_j(x)}, the kernel takes the form
\begin{align}
    \check P(\boldsymbol x) &= \beta\frac{m^2}{(2\pi)^\frac{1}{2}|\vx|}K_1(m|\vx|) \:+ \nonumber\\ &+ \boldsymbol \alpha \cdot \boldsymbol x \left(i\frac{m^2}{(2\pi)^\frac{1}{2}|\vx|^2}K_0(m|\vx|) + i \frac{2m}{(2 \pi)^{\frac{1}{2}}|\vx|^3}K_1(m|\vx|)\right) + \frac{1}{2}\delta^3(\boldsymbol x) \mathds{1}.
\end{align}

\subsection{Proof of Theorem~\ref{thm:1}}

{Let $\psi_n$ be again the sequence described in Section~\ref{sec:construction}.} We compute the expectation value of $\vx^2$ in momentum space
\begin{align}
    \bra{\psi_n}\boldsymbol x^2\ket{\psi_n} &= \langle\nabla_{\boldsymbol p} \widehat\psi_n|\nabla_{\boldsymbol p} \widehat\psi_n\rangle\\
    &=\sum_{j = 1}^3 \int_{\mathbb R ^3}d^3\boldsymbol p \, |\partial_{p_j}\widehat\psi_n|^2.
\end{align}
{Using Eq.~\eqref{eqn: sequence of wave functions in momentum space},} for an arbitrary index $j$, we have,
\begin{align}
    &\int_{\mathbb R ^3}d^3\boldsymbol p  \,|\partial_{p_j}\widehat\psi_n|^2 =\nonumber\\ 
    &= \int_{\mathbb R ^3}d^3\boldsymbol p \, (\partial_{p_j}\widehat{\psi}^*_n)(\partial_{p_j}\widehat\psi_n).\\
    &= \frac{1}{n^3} \int_{\mathbb R^3} d^3\boldsymbol p
    \left(\frac{1}{n}\frac{\partial f^*}{\partial p_j}(\frac{\boldsymbol p}{n})u^\dagger(\boldsymbol p) +
    f^*(\frac{\boldsymbol p}{n})\frac{\partial u^\dagger}{\partial p_j}(\boldsymbol p)\right)
    \left(\frac{1}{n}\frac{\partial f}{\partial p_j}(\frac{\boldsymbol p}{n})u(\boldsymbol p) +
    f(\frac{\boldsymbol p}{n})\frac{\partial u}{\partial p_j}(\boldsymbol p)\right)\\
    &= \underbrace{\frac{1}{n^5} \int_{\mathbb R ^3} d^3\boldsymbol p \frac{\partial f^*}{\partial p_j}(\frac{\boldsymbol p}{n})u^\dagger(\boldsymbol p)\frac{\partial f}{\partial p_j}{(\frac{\boldsymbol{p}}{n})} u(\boldsymbol p)}_{\textit{\textbf{I}}}
    +\underbrace{\frac{1}{n^4}\int_{\mathbb R ^3}d^3\boldsymbol p\frac{\partial f^*}{\partial p_j}(\frac{\boldsymbol p}{n})u^\dagger(p)f(\frac{\boldsymbol p}{n})\frac{\partial u}{\partial p_j}(\boldsymbol p)}_{\textit{\textbf{II}}} \nonumber\\
    &+\underbrace{\frac{1}{n^4}\int_{\mathbb R ^3} d^3\boldsymbol pf^*(\frac{\boldsymbol p}{n})\frac{\partial u^\dagger}{\partial p_j}(\boldsymbol p)\frac{1}{n}\frac{\partial f}{\partial p_j}(\frac{\boldsymbol p}{n})u(\boldsymbol p)}_{\textit{\textbf{III}}}
    + \underbrace{\frac{1}{n^3}\int_{\mathbb R ^3} d^3\boldsymbol pf^*(\frac{\boldsymbol p}{n})\frac{\partial u^\dagger}{\partial p_j}(\boldsymbol p)f(\frac{\boldsymbol p}{n})\frac{\partial u}{\partial p_j}(\boldsymbol p)}_{\textit{\textbf{IV}}}.
\end{align}
We estimate the terms \textit{\textbf{I}}--\textit{\textbf{IV}}. To do this, we use the following Lemma:

\begin{lem}\label{lem:1}
For an arbitrary point $\boldsymbol p\neq 0$ in momentum space, it holds that  
\begin{align}
    \left|\frac{\partial u}{\partial p_j}\right| \leq \frac{2\sqrt{3}}{|\boldsymbol p|}.
    \label{eqn: spinor estimation}
\end{align}
\end{lem}

(A similar estimate was already reported in \cite{Bracken_1999}. We provide a proof in Section~\ref{sec:pfLem1}.)

For \textbf{\textit{I}} we have that
\begin{align}
    \textbf{\textit{I}} &= \frac{1}{n^5} \int_{\mathbb R ^3}d^3\boldsymbol p \: \frac{\partial f^*}{\partial p_j}\Bigl(\frac{\boldsymbol p}{n}\Bigr) \: u^\dagger(\boldsymbol p) \: \frac{\partial f}{\partial p_j}\Bigl(\frac{\boldsymbol p}{n}\Bigr) \: u(\boldsymbol p) \\
    & = \frac{1}{n^5} \int_{\mathbb R ^3} d^3\boldsymbol p \: \frac{\partial f^*}{\partial p_j}\Bigl(\frac{\boldsymbol p}{n}\Bigr) \: \frac{\partial f}{\partial p_j}\Bigl(\frac{\boldsymbol p}{n}\Bigr)\\
    &= \frac{1}{n^2} \int_{\mathbb R ^3}d^3\boldsymbol r \: \frac{\partial f^*}{\partial p_j}(\boldsymbol r) \: \frac{\partial f}{\partial p_j}(\boldsymbol r) \\
    &= \frac{c_1}{n^2}.
\end{align}
In the first step, we used that $u^\dagger u = 1$, and in the second step we substituted $\boldsymbol{r} = \boldsymbol p/n$. The constant $c_1$ is finite because of \eqref{eqn: conditions}. 

Let us turn to \textbf{\textit{II}}.
Suppose $f$ is real valued (or has constant phase). By \eqref{eqn: spinor is normalized},
\begin{align}
 0&=\frac{\partial}{\partial p_j}\Bigl(u^\dagger(\boldsymbol p)  \: u(\boldsymbol{p}) \Bigr)\\ 
 &= \frac{\partial u^\dagger}{\partial p_j}(\boldsymbol p) \: u(\boldsymbol{p}) + u^\dagger(\boldsymbol p) \frac{\partial u}{\partial p_j}(\boldsymbol p)\,.
\end{align}
As a consequence, \textbf{\textit{II}} and \textbf{\textit{III}} cancel each other for such $f$. As an alternative reasoning valid even for general $f$, both \textbf{\textit{II}} and \textbf{\textit{III}} tend to 0 individually:
\begin{align}
    |\textbf{\textit{II}}| 
    &= \left| \frac{1}{n^4}\int_{\mathbb R ^3}d^3\boldsymbol p \: \frac{\partial f^*}{\partial p_j}\Bigl(\frac{\boldsymbol p}{n}\Bigr)  \: u^\dagger(\boldsymbol p) \: f\Bigl(\frac{\boldsymbol p}{n}\Bigr) \: \frac{\partial u}{\partial p_j}(\boldsymbol p) \right|\\
    &\leq  \frac{1}{n^4}\int_{\mathbb R ^3}d^3\boldsymbol p \: \left|\frac{\partial f^*}{\partial p_j}\Bigl(\frac{\boldsymbol p}{n}\Bigr) \: u^\dagger(\boldsymbol p) \: f\Bigl(\frac{\boldsymbol p}{n}\Bigr) \: \frac{\partial u}{\partial p_j}(\boldsymbol p) \right|\\
    &\leq \frac{1}{n^4}\int_{\mathbb R ^3}d^3\boldsymbol p \: \left|\frac{\partial f^*}{\partial p_j}\Bigl(\frac{\boldsymbol p}{n}\Bigr) \: f\Bigl(\frac{\boldsymbol p}{n}\Bigr)\right| \: \left|u(\boldsymbol{p})\right| \: \left|\frac{\partial u}{\partial p_j}(\boldsymbol p)\right|\\
    &\leq \frac{1}{n^4}\int_{\mathbb R ^3}d^3\boldsymbol p \: \left|\frac{\partial f^*}{\partial p_j}\Bigl(\frac{\boldsymbol p}{n}\Bigr) \: f\Bigl(\frac{\boldsymbol p}{n}\Bigr)\right| \: \frac{2\sqrt{3}}{|\boldsymbol p|}\\
    & = \frac{1}{n}\int_{\mathbb R ^3}d^3\boldsymbol r\left|\frac{\partial f^*}{\partial p_j}(\boldsymbol r) \: f(\boldsymbol r)\right| \: \frac{2\sqrt{3}}{n|\boldsymbol r|}\label{eqn: integral 1}\\
    &= \frac{c_2}{n^2}.
\end{align}
In the second step, we applied the Cauchy-Schwarz inequality, in the third \eqref{eqn: spinor estimation}. 
Term \textit{\textbf{III}} is the complex conjugate of \textit{\textbf{II}} and thus also bounded by $c_2/n^2$. Finally, we estimate \textit{\textbf{IV}}:
\begin{align}
    |\textbf{\textit{IV}}| 
    &= \left|\frac{1}{n^3}\int_{\mathbb R ^3} d^3\boldsymbol p \: f^*\Bigl(\frac{\boldsymbol p}{n}\Bigr) \: \frac{\partial u^\dagger}{\partial p_j}(\boldsymbol p) \: f\Bigl(\frac{\boldsymbol p}{n}\Bigr) \: \frac{\partial u}{\partial p_j}(\boldsymbol p)\right| \\
    &\leq \frac{1}{n^3} \int_{\mathbb R ^3} d^3\boldsymbol p \: \left|f^*\Bigl(\frac{\boldsymbol p}{n}\Bigr)\right| \: \frac{2\sqrt{3}}{|\boldsymbol p|} \: \left|f\Bigl(\frac{\boldsymbol p}{n}\Bigr)\right| \: \frac{2\sqrt{3}}{|\boldsymbol p|}\\
    &= \frac{1}{n^2}\int_{\mathbb R ^3} d^3\boldsymbol r \: \frac{12|f(\boldsymbol r)|^2}{\boldsymbol r^2}\\
    &= \frac{c_3}{n^2} \,.
\end{align}
In the first step, we used \eqref{eqn: spinor estimation} again. Putting the estimates together, We find that 
\begin{align}
    \lim_{n\to\infty} \sigma^2_{\psi_n} &= \lim_{n\to\infty} \bra{\psi_n}\boldsymbol x^2\ket{\psi_n} - \bra{\psi_n}\boldsymbol x\ket{\psi_n}^2\\ &\leq \lim_{n\to\infty}\bra{\psi_n}\boldsymbol x^2\ket{\psi_n}\\ &=0.
\end{align}

\subsection{Proof of Theorem~\ref{thm:2}}

Let $d\geq 1$ (the dimension of $\vx$) and $n\geq 2$ (the index of the sequence $\rho_n$), $\ve_1=(1,0,\ldots,0) \in \mathbb R^d$ the first canonical basis vector, and set
\begin{align}
    g(\vx)&:= \pi^{-d/2}e^{-\vx^2}\\
    \rho_n(\vx) &:= (n-2)n^{d-1} g(n\vx) + \frac{1}{n}g(\vx-n\ve_1)+ \frac{1}{n}g(\vx+n\ve_1).
\end{align}
In the following, we write
\begin{align}
    ||g(\vx)||_1 = \int d^d\vx \, |g(\vx)|,
\end{align}
first in order to distinguish the $L^1$ from the $L^2$ norm, and second in a slight abuse of notation (writing $g(\vx)$ instead of $g$ for the function). Each $\rho_n$ is normalized as
\begin{align}
    ||\rho_n(\vx)||_1 
    &= (n-2)n^{d-1}||g(n\vx)||_1 + \frac{1}{n}||g(\vx-n\ve_1)||_1+ \frac{1}{n}||g(\vx+n\ve_1)||_1\\
    &=\frac{n-2}{n}||g(\vx)||_1 + \frac{1}{n}||g(\vx)||_1+ \frac{1}{n}||g(\vx)||_1\\
    &= ||g(\vx)||_1\\
    &= 1.
\end{align}
To prove property{~\eqref{eqn: delta property}}, we note that
\begin{align}
    \rho_n(\vx)-{\frac{n-2}{n}}n^d g(n\vx) = 
    \frac{1}{n}\Bigl(g(\vx-n\ve_1)+g(\vx+n\ve_1) \Bigr)  \,.
\end{align}
{Thus, for an arbitrary Schwartz function $h(\vx)$, 
\begin{equation}
    \biggl| \int_{\mathbb{R}^d} d^d\vx\Bigl(\rho_n(\vx)-\frac{n-2}{n}n^dg(n\vx)\Bigr)h(\vx) \biggr| \leq \frac{2}{n}\sup_{\vx} |h(\vx)|
\end{equation}
tends to 0 as $n\to\infty$.
Since the delta distribution can be defined by the sequence $n^dg(n\vx)$,
\begin{align}
    \lim_{n\to\infty} \int_{\mathbb{R}^d} d^d\vx\, \rho_n(\vx)\, h(\vx) 
    &= \lim_{n\to\infty} \frac{n-2}{n} \int_{\mathbb{R}^d} d^d\vx\, n^dg(n\vx)\, h(\vx) \\
    &= \lim_{n\to\infty}  \int_{\mathbb{R}^d} d^d\vx\, n^dg(n\vx)\, h(\vx) \\
    &= h(\boldsymbol{0})\,,    
\end{align}
so}
\begin{align}
    \lim_{n\to\infty} \rho_n(\vx) = \delta^d(\vx).
\end{align}

Property~\eqref{eqn: infinite variance} follows from
\begin{align}
    &\int_{\mathbb{R}^d}d^d\vx \, \rho_n(\vx) \, \vx^2  \nonumber\\
    &= \int_{\mathbb{R}^d} d^d\vx \Bigl( (n-2)n^{d-1}g(n\vx) \, \vx^2 + \frac{1}{n}g(\vx-n\ve_1) \, \vx^2+\frac{1}{n}g(\vx+n\ve_1) \, \vx^2 \Bigr)\\
    &\geq \int_{\mathbb{R}^d}d^d\vx \, \frac{1}{n}g(\vx-n\ve_1) \, \vx^2\\
    &\geq \int_{[n,\infty)}dx_1 \, \frac{1}{n} \pi^{-1/2} \exp(-(x_1-n)^2) \, x_1^2\\
    &\geq \int_{[n,\infty)}dx_1 \, \frac{1}{n} \pi^{-1/2} \exp(-(x_1-n)^2) \, n^2\\
    &= \frac{n}{2}.
\end{align}
Finally, it is easy to see that \eqref{eqn: zero expectation value} holds, as $\rho_n(\vx)$ is even in each $x_j$.

\subsection{Proof of Remark~\ref{rem:converse}}

Let
\begin{align}
    \rho(x) \geq 0,\\
    \int_\mathbb{R} dx \: \rho_n(x) = 1,\\
    \lim_{n\to\infty}\int_\mathbb{R} dx \: x^2 \, \rho_n(x) = 0.
    \label{eqn: limit x^2}
\end{align}
We want to prove that 
\begin{align}
    \lim_{n\to\infty}\rho_n(x) = \delta(x).
\end{align}
To do this, we need to show that for any test function $f$, we have
\begin{align}
    \lim_{n\to\infty}\int_\mathbb R dx \: \rho_n(x) \: f(x) = f(0).
    \label{eqn: delta property 2}
\end{align}
{For an arbitrary $r>0$} we have that
\begin{align}
    \int_{\mathbb R \setminus B_{{r}}(0)}dx \: x^2 \, \rho_n(x) \geq {{r^2}} \int_{\mathbb R \setminus B_{{r}}(0)}dx \: \rho(x),
\end{align}
which by \eqref{eqn: limit x^2} implies that
\begin{align}
    \lim_{n\to\infty} \int_{\mathbb{R}\setminus B_{{r}}(0)} dx \: \rho_n(x) = 0,\\
    \lim_{n\to\infty} \int_{B_{{r}}(0)} dx \: \rho_n(x) = 1.
\end{align}
Now, let $\varepsilon > 0$ and {$r>0$ be} such that
\begin{align}
    |f(x)-f(0)| < \varepsilon \: \forall x \in B_r(0).
\end{align}
We estimate
\begin{align}
    &\left|\int_\mathbb R dx \; \rho_n(x)f(x) - f(0)\right| = \left|\int_\mathbb R dx \: \rho_n(x) \, (f(x) - f(0)) \right|\\
    &\leq \int_{B_{{r}}(0)} dx\: \rho_n(x) \, |f(x)-f(0)| + \int_{\mathbb R \setminus B_{{r}}(0)} dx\: \rho_n(x) \, |f(x)-f(0)|\\
    &< \varepsilon\int_{B_{{r}}(0)} dx \: \rho_n(x) + \max_{x\in\mathbb R}|f(x)-f(0)|\int_{\mathbb R \setminus B_{{r}}(0)} dx \:\rho_n(x)\,, 
\end{align}
and thus we have that
\begin{align}
     {\limsup_{n\to\infty}}\left|\int_\mathbb R dx \, \rho_n(x) \, f(x) - f(0)\right| {\leq} \varepsilon.
\end{align}
Finally, $\varepsilon$ was arbitrary, so \eqref{eqn: delta property 2} is established.

\subsection{Proof of Lemma~\ref{lem:1}}
\label{sec:pfLem1}

In the proof, we make use of the following easy inequalities:
\begin{align}
    \max\{|\boldsymbol p|,m\} &\leq E(\boldsymbol p)\leq E(\boldsymbol p)+m,\\
    2m|\boldsymbol p|&\leq m^2 + \boldsymbol p^2.
\end{align}
First we simplify $u_1(\boldsymbol p)$:
\begin{align}
    u_1(\boldsymbol p) &= \frac{E(\boldsymbol p)+m}{\sqrt{2E(\boldsymbol p)(E(\boldsymbol p)+m)}}\\
    &= \sqrt{\frac{(E(\boldsymbol p)+m)^2}{2E(\boldsymbol p)(E(\boldsymbol p)+m)}}\\
    &= \sqrt{\frac{E(\boldsymbol p)+m}{2E(\boldsymbol p)}}\\
    &= \sqrt{\frac{1}{2}+ \frac{m}{2E(\boldsymbol p)}}.
\end{align}
Differentiating then gives
\begin{align}
    \left|\frac{\partial u_1}{\partial p_k}\right| 
    &= \left|\frac{1}{2\sqrt{\frac{1}{2}+ \frac{m}{2E(\boldsymbol p)}}} \frac{\partial}{\partial p_k}\left(\frac{1}{2}+ \frac{m}{2E(\boldsymbol p)}\right)\right|\\
    &= \left|\frac{1}{2\sqrt{\frac{1}{2}+ \frac{m}{2E(\boldsymbol p)}}} \frac{m}{4(\boldsymbol p^2+m^2)^\frac{3}{2}}2p_k \right|\\
    &\leq \left|\frac{1}{\sqrt{2}} \frac{m|\boldsymbol p|}{2(\boldsymbol p^2+m^2)^\frac{3}{2}} \right|\\
    &\leq \left|\frac{1}{\sqrt{2}} \frac{\boldsymbol p^2+m^2}{4(\boldsymbol p^2+m^2)^\frac{3}{2}} \right|\\
    &\leq \left|\frac{1}{\sqrt{2}} \frac{1}{4(\boldsymbol p^2+m^2)^\frac{1}{2}} \right|\\
    &\leq \frac{1}{4\sqrt{2}}\frac{1}{|\boldsymbol p|}.
\end{align}
The second component is trivial,
\begin{align}
    \left|\frac{\partial u_2}{\partial p_k}\right| = 0.
\end{align}

\begin{align}
    \left|\frac{\partial u_3}{\partial p_k}\right| &= \left|\frac{\partial}{\partial p_k}\frac{p_3}{\sqrt{2E(\boldsymbol p)(E(\boldsymbol p)+m)}}\right|\\
    &= \left|\frac{\delta_{3k}}{\sqrt{2E(\boldsymbol p)(E(\boldsymbol p)+m)}}-\frac{p_3}{2(2E(\boldsymbol p)(E(\boldsymbol p)+m))^{\frac{3}{2}}} \frac{\partial}{\partial p_k}\left(E(\boldsymbol p)(E(\boldsymbol p)+m)\right)\right|\\
    &= \left|\frac{\delta_{3k}}{\sqrt{2E(\boldsymbol p)(E(\boldsymbol p)+m)}}-\frac{p_3}{2(2E(\boldsymbol p)(E(\boldsymbol p)+m))^{\frac{3}{2}}} \left(\frac{p_k}{E(\boldsymbol p)}(E(\boldsymbol p)+m)+p_k\right)\right|\\
    &= \left|\frac{\delta_{3k}}{\sqrt{2E(\boldsymbol p)(E(\boldsymbol p)+m)}}-\frac{p_3p_k}{2^\frac{5}{2}E(\boldsymbol p)^\frac{5}{2}(E(\boldsymbol p)+m)^{\frac{1}{2}}}-\frac{p_3p_k}{2(2E(\boldsymbol p)(E(\boldsymbol p)+m))^{\frac{3}{2}}} \right|\\
    &\leq \left|\frac{\delta_{3k}}{\sqrt{2E(\boldsymbol p)(E(\boldsymbol p)+m)}} \right|+\left|\frac{p_3p_k}{2^\frac{5}{2}E(\boldsymbol p)^\frac{5}{2}(E(\boldsymbol p)+m)^{\frac{1}{2}}}\right|+\left|\frac{p_3p_k}{2(2E(\boldsymbol p)(E(\boldsymbol p)+m))^{\frac{3}{2}}} \right|\\
    &\leq \frac{1}{\sqrt{2}|\boldsymbol p|} +\frac{|\boldsymbol p|^2}{2^\frac{5}{2}|\boldsymbol p|^\frac{5}{2}|\boldsymbol p|^{\frac{1}{2}}}+\frac{|\boldsymbol p|^2}{2(2|\boldsymbol p|^2)^{\frac{3}{2}}}\\
    &= \frac{3}{2\sqrt{2}|\boldsymbol p|}
\end{align}
Analogous steps also hold for the fourth component,
\begin{align}
    \left|\frac{\partial u_4}{\partial p_k}\right| &\leq \left|\frac{\delta_{1k}-i\delta_{2k}}{\sqrt{2E(\boldsymbol p)(E(\boldsymbol p)+m)}} \right|+\left|\frac{(p_1-ip_2)p_k}{2\cdot2^\frac{3}{2}E(\boldsymbol p)^\frac{5}{2}(E(\boldsymbol p)+m)^{\frac{1}{2}}}\right|+\left|\frac{(p_1-ip_2)p_k}{2(2E(\boldsymbol p)(E(\boldsymbol p)+m))^{\frac{3}{2}}} \right|\\
    &\leq \left|\frac{1}{\sqrt{2}|\boldsymbol p|} \right|+\left|\frac{(p_1-ip_2)p_k}{4\sqrt{2}|\boldsymbol p|^3}\right|+\left|\frac{(p_1-ip_2)p_k}{4\sqrt{2}|\boldsymbol p|^3} \right|\\
    &\leq \frac{\sqrt{2}}{|\boldsymbol p|}.
\end{align}
This completes the proof of Lemma~\ref{lem:1}.

\section{Related Works in the Literature}
\label{sec:literature}


{An alternative proof of Theorem~\ref{thm:1} can presumably be obtained from a different sequence $\tilde\psi_n$ of Dirac wave functions that were introduced in \cite{BB19}; there, a wave function $\Psi_+ \in\Hilbert_+$ was defined in (4) and (9) (equation numbers now referring to \cite{BB19}) depending on a parameter $a>0$ with the dimension of a length, and $\tilde\psi_n=\Psi_+$ with $a=\lambda_C/n$. It is given in (9) in the form $\Psi_+=N \phi$, where $N$ is a normalization factor and $\phi$ an expression that converges for every $\vx\neq \vzero$ to a function $\phi_0$ as $a\to 0$. For any $a$, $|\Psi_+(\vx)|^2$ is an even function of $\vx$, so $\langle \vx \rangle := \int_{\RRR^3}d^3\vx \, \vx \, |\Psi_+(\vx)|^2=0$. From (10) it follows (considering only $t=0$) that $\int_{\RRR^3}d^3\vx \, \vx^2 \, |\phi_0(\vx)|^2 < \infty$, and from (12) that $N\to 0$ as $a\to 0$; this suggests that $\langle \vx^2\rangle = \int_{\RRR^3} d^3\vx \, \vx^2 \, |\Psi_+(\vx)|^2$ tends to 0 as $a\to 0$.

For a massless particle, $m=0$, there is a very simple scaling argument \cite{BB22} {proving} that the position uncertainty can be arbitrarily small: For any $0<s<\infty$, consider the unitary scaling operator $R_s:\Hilbert \to \Hilbert$ defined by
\be\label{Rsdef}
(R_s\psi)(\vx):= s^{3/2} \, \psi(s \vx)\,.
\ee
For $m=0$, $H_D$ is scale {covariant in the sense that}
\be
R_{1/s} \, H_D \, R_s = {s} \, H_D \,.
\ee
Therefore, if $\psi \in \Hilbert_+$ with $\|\psi\|=1$, also $R_s \psi$ lies in $\Hilbert_+$ (and has norm 1) for any $0<s<\infty$. But the spread of $R_s \psi$ differs from that of $\psi$ by a factor $s^{-1}$,
\be
\sigma_{R_s \psi} = s^{-1} \, \sigma_\psi\,.
\ee
Therefore, there is no positive lower bound to $\sigma_\psi$ for $\psi\in\Hilbert_+$. 

{Also for} $m>0$, Bialynicki-Birula and Bialynicka-Birula \cite{BB22} sketched an argument} {against any positive lower bound to $\sigma_\psi$; we describe here our understanding of it. Let $H_m$ be $H_D$ for a given value of $m\geq 0$ and $\Hilbert_+(m)$ the positive spectral subspace of $H_m$. Then, as one easily checks, 
\be
R_{1/s} \, H_m \, R_s = s \, H_{m/s} \,.
\ee
Therefore, 
\be
\text{if }\psi \in \Hilbert_+(m),\text{ then }R_s\psi \in \Hilbert_+(sm) 
\ee
(and has width $s^{-1} \sigma_\psi$). 
In order to find $\psi\in\Hilbert_+(m)$ with $\|\psi\|=1$ and $\sigma_\psi<a$ for given (arbitrarily small) length $a$, find first $\phi\in\Hilbert_+(0)$ with $\|\phi\|=1$ and $\sigma_\phi < a$ (which was shown above to exist); suppose $\varepsilon>0$ is small; since $\Hilbert_+(\varepsilon m)$ lies close to $\Hilbert_+(0)$, there should be $\chi\in \Hilbert_+(\varepsilon m)$ close to $\phi$ with $\|\chi\|=1$; since $\chi$ is close to $\phi$, it should still have $\sigma_\chi < 2 a$; set $\psi=R_{1/\varepsilon} \chi$, so $\psi\in\Hilbert_+(m)$ and $\sigma_\psi= \varepsilon \sigma_\chi< 2\varepsilon a<a$.

The simplest strategy for obtaining an alternative mathematical proof of Theorem~\ref{thm:1} in this way starts perhaps by considering the first vector $\psi=\psi_1$ in Bracken and Melloy's sequence for Gaussian $f$, i.e., 
\be
\psi(\vx) = \psi_{\ell,m} (\vx) = \mathcal{N}_{\ell,m} \int_{\RRR^3} d^3\vp \, e^{-\ell^2 \vp^2} \, u_m(\vp) \, e^{i\vx\cdot \vp}
\ee
with $\mathcal{N}_{\ell,m}$ the normalizing constant, $\ell$ the parameter of the Gaussian, and $u_m(\vp)$ the appropriate eigenspinor for $m\geq 0$, given by \eqref{udef} and \eqref{Edef} for $m>0$ and by
\be
u_0(\vp)=\frac{(|\vp|,0,p_3,(p_1-ip_2))}{\sqrt{2}|\vp|}
\ee
for $m=0$. Note that $R_s \psi_{\ell,m} = \psi_{\ell/s,sm}$. Take $\phi=\psi_{\ell,0}$ for sufficiently small $\ell$, then $\chi=\psi_{\ell,\varepsilon m}$ for sufficiently small $\varepsilon$, prove that $\sigma_\chi<2a$, and apply $R_{1/\varepsilon}$. 
}


We also mention a similar result of Dodonov and Mizrahi \cite{DM93}, which showed that for any $j\in\{1,2,3\}$ and the modified position operator $X_j=P_+ x_j P_+ + P_- x_j P_-$ with $P_\pm$ the projection to $\Hilbert_\pm$, the variance can be arbitrarily close to 0 for unit vectors in $\Hilbert_+$. However, this fact does not imply that the variance of $x_j$ can be arbitrarily close to 0 for such vectors.\footnote{But conversely, Theorem~\ref{thm:1} implies their result because, for $\psi\in \Hilbert_+$ with $\langle \psi|x_j|\psi\rangle =0$ for all $j$ (so $\langle \psi|X_j|\psi\rangle=0$), $\langle \psi|\vx^2|\psi\rangle \geq \langle \psi|x^2_j|\psi\rangle = \langle \psi|(P_+x_jP_+x_jP_+)+(P_+x_jP_-x_jP_+)|\psi\rangle \geq \langle \psi|P_+x_jP_+x_jP_+|\psi\rangle = \langle \psi|X_j^2|\psi\rangle$. On the other hand, their proof is simpler than the proof of Theorem~\ref{thm:1}.}

\bigskip

\noindent{\bf Acknowledgments.} We thank Iwo Bialynicki-Birula and Viktor Dodonov for helpful discussion.

\noindent{\bf Funding.} This work received no external funding. 

\noindent{\bf Conflict of interest.} The authors declare no conflict of interest. 

\noindent{\bf Data availability.} No data were generated or collected in this work.

\bibliographystyle{unsrt} 
\bibliography{references.bib} 

\end{document}